\documentclass{aa}
\usepackage{txfonts}
\usepackage{natbib}
\bibpunct{(}{)}{;}{a}{}{,} % to follow the A&A style
\usepackage{graphicx}

\title{Dust from Reionization}
\author{Erik Elfgren \inst{1}
	\and Fran\c cois-Xavier D\'esert \inst{2}}
\institute{Department of Physics, Lule\aa\ University of Technology, SE-971 87 Lule\aa, Sweden
		\and Laboratoire d'Astrophysique, Observatoire de Grenoble,
		     BP 53, 414 rue de la piscine, 38041 Grenoble Cedex 9, France}
\offprints{Erik Elfgren,
\email{elf@ludd.luth.se}}

\date{Received <date> / Accepted <date>}

\begin{document}

\abstract{
The possibility that population III stars
have reionized the Universe at redshifts greater than 6 has recently
gained momentum with WMAP polarization results. Here we analyse the
role of early dust produced by these stars and ejected into the
intergalactic medium.  We show that this dust, heated by the
radiation from the same population III stars,
produces a submillimetre excess.
The electromagnetic spectrum of this excess could account for a significant fraction of the FIRAS
(Far Infrared Absolute Spectrophotometer) cosmic far infrared background above 700 micron.
This spectrum, a primary anisotropy ($\Delta T$) spectrum times the $\nu^2$
dust emissivity law, peaking in the submillimetre domain around 750 micron,
is generic and does not depend on other detailed dust
properties.  Arcminute--scale anisotropies, coming from inhomogeneities in this
early dust, could be detected by future submillimetre
experiments such as Planck HFI.

}
\maketitle

\section{Introduction}
\begin{figure}%[here!]
	\resizebox{\hsize}{!}{\includegraphics{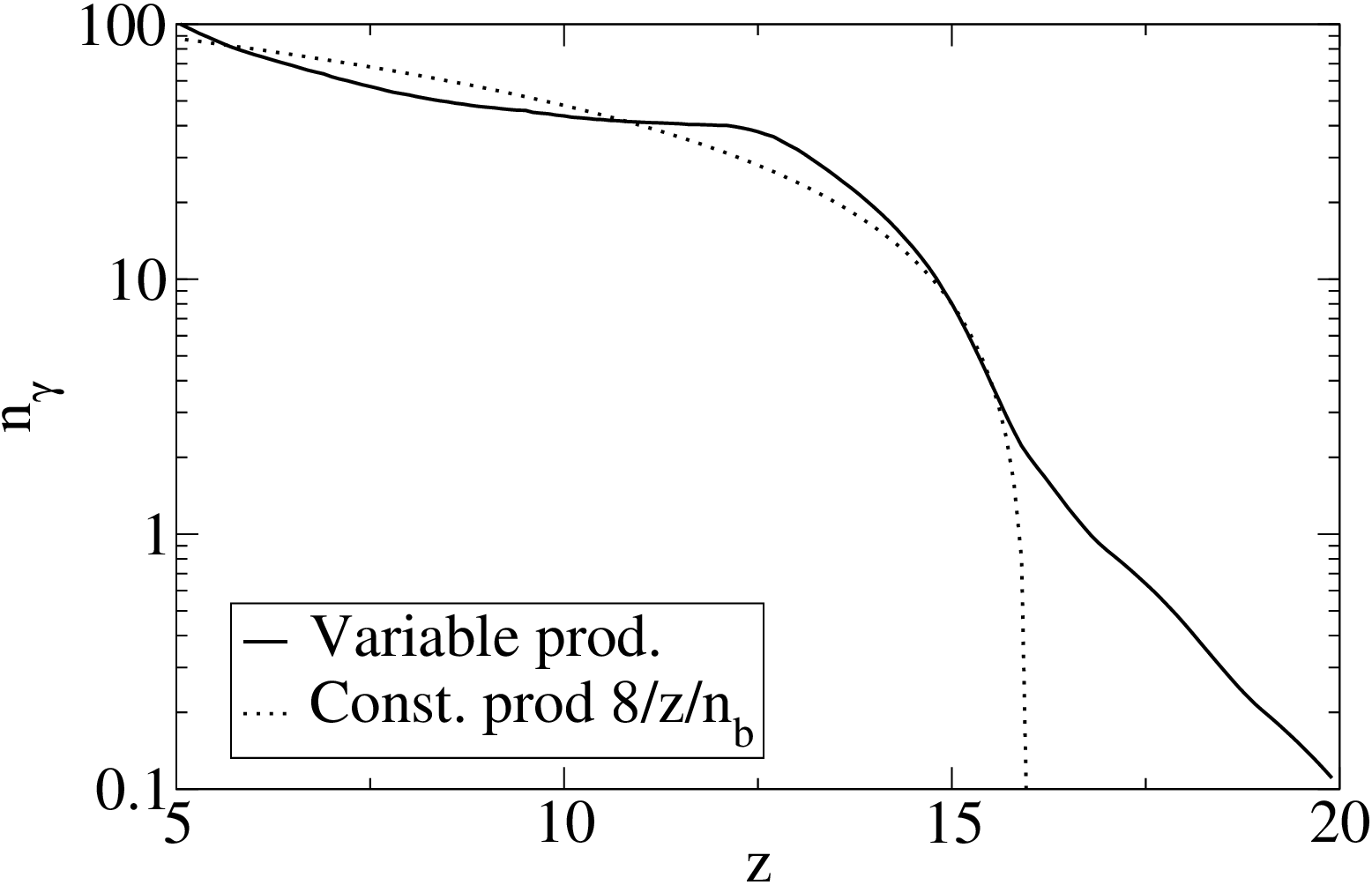}}
	\caption{Total number of ionizing photons produced from Pop III stars per baryon,
	cf. \cite[ figure 14]{Cen:2002zc}. The dotted line represents a simplified model
	with a constant photon production, from $z=16$, of 8 per unit $z$ per baryon. The results are similar.}
	\label{fig:ng}
\end{figure}

\begin{figure}%[here!]
	\resizebox{\hsize}{!}{\includegraphics{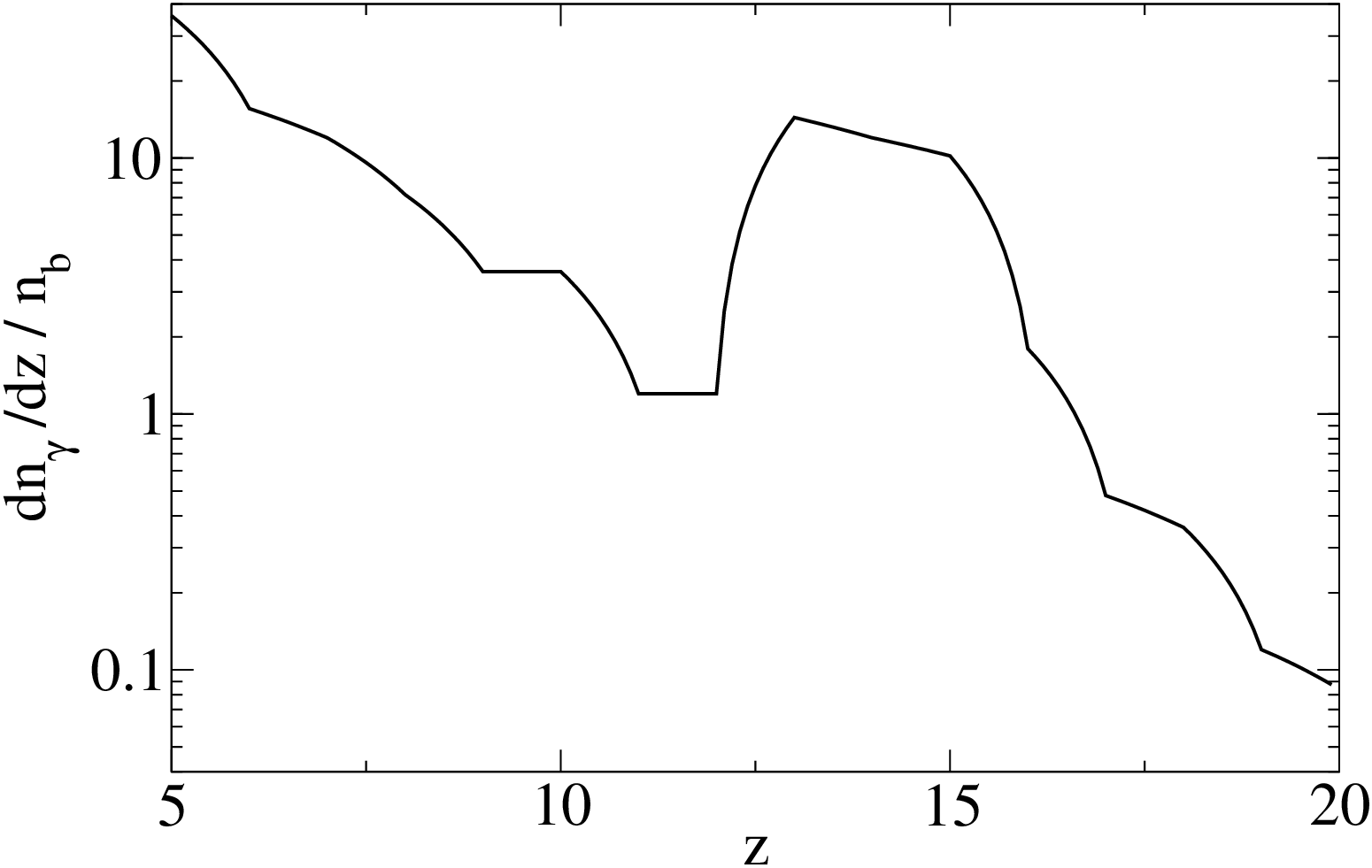}}
	\caption{Production rate of ionizing photons from Pop III stars per baryon, $\frac{dn_\gamma}{dz}/n_b$.
		The odd form between each integer $z$ is not physical
		but are due to the fact that the redshift $z$ is a nonlinear function of
		time.}
	\label{fig:dng_dz}
\end{figure}

More accurate measurements of the cosmic microwave background (CMB) implies a
need for a better understanding of the different foregrounds.
We study the impact of dust in the very early universe $5<z<15$.
WMAP data on the CMB polarization, \cite{2003ApJS..148..161K} provides a strong
evidence for a rather large Thomson opacity during the reionization of
the Universe: $\tau_e=0.17\pm0.04\, (68\% C.L.)$. Although the mechanism of
producing such an opacity is not fully understood, \cite{Cen:2002zc,Cen:2003ey}
has shown that 
early, massive population--III (Pop III) stars could ionize the Universe within
$5<z<15$ (see Fig.~\ref{fig:ng} and Fig.~\ref{fig:dng_dz}).
Adopting this hypothesis, we discuss the
role and observability of the dust that is produced by the Pop III
stars. As we can only conjecture about the physical properties and
the abundance of this early dust, we adopt a simple dust grain model with
parameters deduced from the Milky Way situation. The dust production
is simply linked to the ionizing photon production by the stars through
their thermal nuclear reactions.
The low potential well of the small pre-galactic halos allows the ejected dust
to be widely spread in the intergalactic medium.
The ionizing and visible photons from the
same Pop III stars heat this dust.
There are no direct measurements of this dust, but by means of other results the amount
of dust can be estimated. A similar study has been done for a later epoch of the
universe, in which data are more readily available, \cite{1999ApJ...522..604P}.
We use a cosmology with $\Omega_{tot} = \Omega_m + \Omega_\Lambda = 1$, where
$\Omega_m = \Omega_b + \Omega_{DM} = 0.133/h^2$, $\Omega_b = 0.0226/h^2$ and
$h = 0.72$ as advocated by WMAP, \cite{2003ApJS..148..175S},
using WMAP data in combination with large scale structure observations %Myref 80p15t7.
(2dFGRS + Lyman $\alpha$). 
Furthermore, since $z\gg1$ the universe is matter-dominated.
We relate all cosmological parameters to their measurement today so that
they have their present-day values throughout our calculations.

We now proceed to
compute the abundance and the temperature of this dust. Consequences on
the CMB distortions are then to be discussed.


\section{Dust Model}
\label{sec:DustEv}
Here we assume the dust properties to be similar to what is observed in our galaxy.
For simplicity, 
we suppose spherical dust grains with radius $a=0.1\mu$m and density
$\rho_g = 2.3 \cdot 10^3$ kg/m$^3$.% (see e.g. \cite{1999A&A...344..322L}).
The absorption cross section, $\sigma_\nu$,
between photons and dust can be written as
\begin{equation}
	\sigma_{\nu}=Q_\nu\pi a^{2},
\label{eq:sigma}
\end{equation}
where we parametrize the frequency dependency as
\begin{equation}
	Q_\nu=\left\{\begin{array}{ll}
		Q_0 \frac{a}{a_r}\left(\frac{\nu}{\nu_r}\right)^{\beta_\nu} &
		\textrm{submm and infra red (IR),}\\
		1 &
		\textrm{visible and ultra violet (UV),}\\
	\end{array}\right.
\end{equation}
where %$a$ is the grain size,
$\nu_r, a_r$ and $Q_0$ are normalization constants. There is only one independent
constant which means that we can fix $a_r = 0.1\,\mu$m. In
\cite[ figure 3]{1990A&A...237..215D} the poorly known knee wavelength, $\lambda_r=c/\nu_r$ was set to 100
$\mu$m. Here, we choose 40 $\mu$m for simplicity, so that early dust radiates mostly in the
$\nu^2$ emissivity regime. %$\nu_r = c/(40\,\mu$m). %XAVIER REF?
Above the characteristic frequency $\nu_r$ the spectral index $\beta=1$,
below $\beta=2$. The exact position of $\nu_r$ is not very important for
our study because it is mainly above the interesting wave-length region $\sim 0.3 - 3$ mm
and it will not change the magnitude of the signal.

In the submm and far IR range, the spectral index is constant, and with 
$Q_0 = 0.0088$ the assumed opacity agrees well with measurements by
FIRAS on cirrus clouds in our galaxy, cf. \cite{1996A&A...312..256B,1998ApJ...508..123F,1999A&A...344..322L}.
In the visible and UV region, the cross section is independent of the frequency
because $\lambda<2\pi a$. In the submm region, the cross section is proportional
to the mass of the grain.

In order to evaluate the significance of the dust during the reionization, we calculate the amount
of dust present in the universe at a given time.
The co-moving relative dust density is
$\Omega_{d,0} = \rho_d(z)/((1+z)^{3}\rho_c)$,
where $\rho_d(z)$ is the dust density, $z$ is the red-shift, $\rho_c = \frac{3H_0^2}{8\pi G}$ is the critical density
($H_0$ and $G$ are Hubble's and Newton's constants, respectively).
The co-moving relative dust density as measured today evolves as:
\begin{equation}
        \frac{d\Omega_{d,0}}{dz} = J_+ - J_-,
\label{eq:DustEv}
\end{equation}
where $J_+$ and $J_-$ are the production and the destruction rate respectively.

The Pop III stars produce enough photons for the reionization while
burning $H$ and thus forming metals ($Li$ and higher). These metals are released in
supernovae explosions at the end of the stars short lives ($\sim 1$ Myr),
whereafter they clump together to form dust, \cite{2003astro.ph..7108N}.
Knowing the production rate of ionizing photons to be $\frac{dn_\gamma}{dz}/n_b$
(Fig.~\ref{fig:dng_dz}), we can calculate the total photon energy released from the Pop III stars.
This can be done by supposing that each photon has an effective energy of
$E_\gamma = c_\gamma\int_{\nu_{ion}}^\infty d\nu\, h\nu B_\nu(T_*)/\int_{\nu_{ion}}^\infty d\nu\, B_\nu(T_*),$
where $h\nu_{ion} = 13.6$ eV and $B_\nu(T_*)$ is the spectrum of a star with temperature $T_*$.
The energy of the non-ionizing photons is included through $c_\gamma = u_{tot}/u_{\nu>\nu_{ion}}$
($u$ is the energy from the star). % with temperature $T_*$).
A Pop III star has $T_*\sim$ 80~000 K \cite[ page 9]{2002ApJ...576...36S} which gives 
$E_\gamma \approx 36$ eV.
Note that for other reasonable star temperatures, 
$E_\gamma$ does not vary significantly,
$E_\gamma|_{60\times 10^3 K}\approx 36$ eV and 
$E_\gamma|_{100\times 10^3 K}\approx 40$ eV.
Hence, the total Pop III photon energy production is
$E_\gamma \frac{dn_\gamma}{dz}/n_b$ per baryon per unit $z$. %\approx \frac{dn_\gamma}{dz}E_\gamma$.
For each consumed nucleon, we assume that a nuclear energy of $E_r = 7$ MeV is released as radiation,
which means that the nucleon consumption rate is
$ \frac{E_\gamma}{E_r} \frac{dn_\gamma}{dz}/n_b$ nucleons per baryon per unit $z$.
If $f_d$ is the fraction of the consumed baryon mass that becomes interstellar dust,
(some of the metal atoms will remain with the core after the SN explosion,
some will stay in the close vicinity of the SN and some will never clump together to form dust)
the co-moving dust production rate will be
\begin{equation}
        J_+ = f_d \frac{E_\gamma}{E_r}\Omega_b\frac{dn_\gamma}{dz}/n_b.
\end{equation}

A dust grain will eventually be destroyed, e.g. by collision, by supernova shockwaves
or by cosmic rays, see \cite{1979ApJ...231..438D} for further discussion.
If a dust grain has a lifetime of $\Delta t$ we can write
the dust destruction rate as
\begin{equation}
        J_- = \frac{\Omega_{d,0}(z)}{\Delta t}\frac{dt}{dz} \approx
	-\frac{\Omega_{d,0}(z)}{\Delta tH_0\Omega_m^{1/2}(1+z)^{5/2}},
\end{equation}
where $\Omega_m$ is the relative matter content today,
because the universe is matter dominated for $5<z<15$.

Solving Eq.~\ref{eq:DustEv} gives the dust density evolution
\begin{equation}
        \Omega_{d,0}(z) = \int_z^{z_i} J_+(z')\frac{Y(z')}{Y(z)}dz',
\end{equation}
where $z_i=20$ is the beginning of the dust formation (see Fig.~\ref{fig:ng}) and 
\begin{equation}
        Y(z) = %e^{\int_z^\infty Pdz'} =
        %e^{\int_z^\infty \frac{(z'+1)^{-5/2}}{\Omega_m^{1/2}\Delta t H_0}dz'} =
        exp \left(\frac 23\frac{(1+z)^{-3/2}}{\Omega_m^{1/2}\Delta t H_0}\right).
\end{equation}
We note that the source term $J_+$ is modulated by the destruction term
$\frac{Y(z')}{Y(z)}$.
        %C = e^{\frac 32\frac{y^{-3/2}}{\Delta t H_0}}\Omega_{d,0}(y_i) = 0.
%
\begin{figure}%[here!]
	\resizebox{\hsize}{!}{\includegraphics{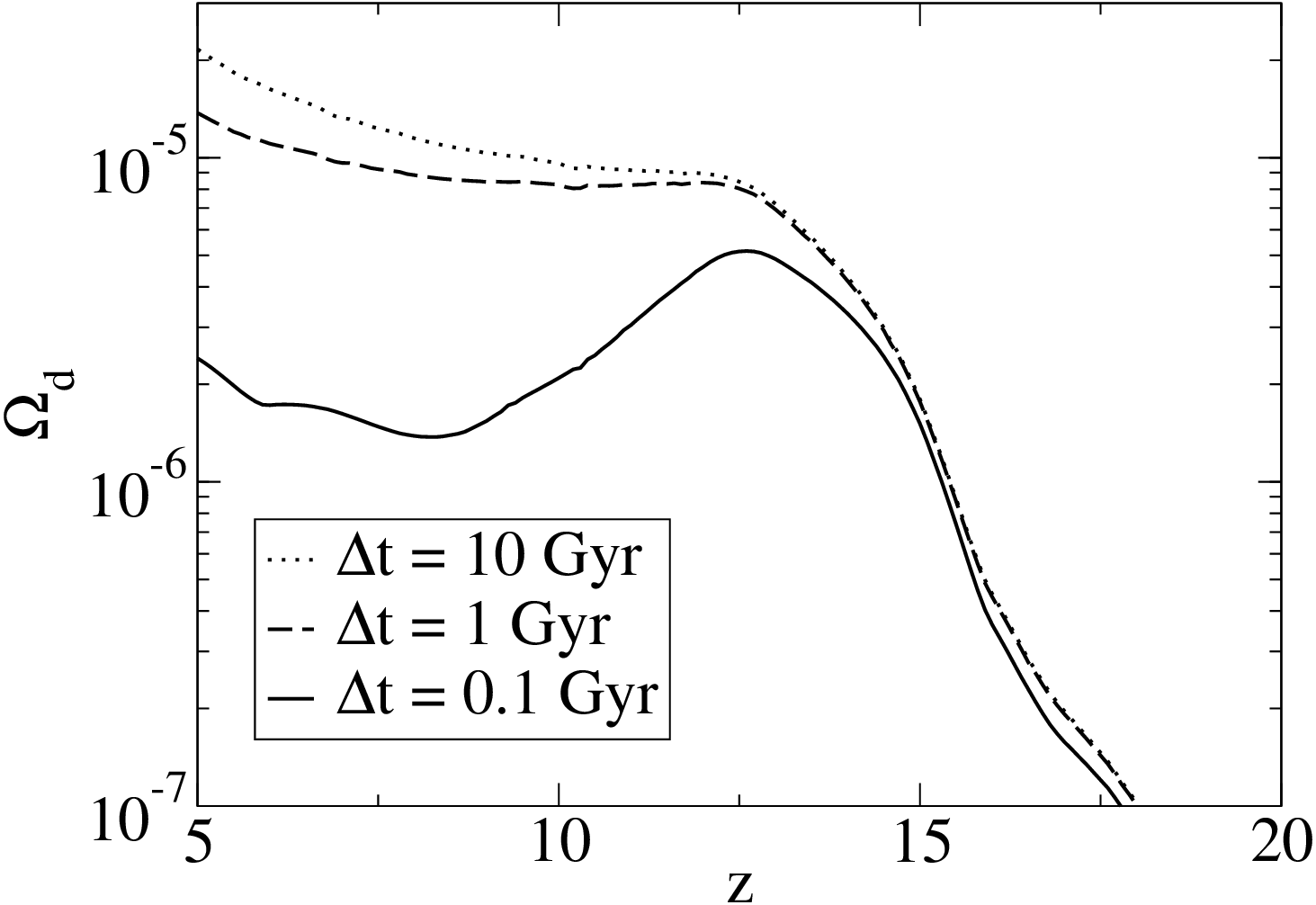}}
\caption{The co-moving relative dust density evolution $\Omega_{d,0} = \rho_{dust}/\rho_c$, for $f_d=1$.
        %Dashed lines are calculated with a constant photon production $8/z/n_b$, solid lines are
        %calculated with a variable photon production according to figure \ref{fig:ng}.
	The minima at $z=6$ and 9 for $\Delta t \le 0.1$ Gyr is due to the fact that
	$\Delta z = 1$ is not a constant time interval.}
\label{fig:rho_d_ev}
\end{figure}
The dust density is plotted in Fig.~\ref{fig:rho_d_ev} where we note
a strong dependency on the dust lifetime.
In local dust $\Delta t\sim 100$ Myr, \cite{1979ApJ...231..438D}.
However, the uncertainty is rather large, % and strongly dependent on the environment.
according to \cite{1990eism.conf..193D},
$\Delta t = 30$ Myr -- $10$ Gyr,
depending on the environment. 
Note, however, that the density at the reionization red-shifts
is much lower than in the interstellar medium in the Milky Way
which implies a rather long dust life-time.

\section{Results and Discussion}
\subsection{Metallicity}
\label{sec:Met}
If we suppose that most of the metals were ejected as dust (not as gas)
the metallicity comes from the dust grains. 
The metallicity is directly obtained through the produced dust. By letting
$\Delta t \rightarrow \infty$ ($\Delta t = 10$ Gyr is good enough) we find the metallicity:
\begin{equation}
        \frac{Z}{Z_\odot} = %\frac{\Omega_{d,0}(\Delta t \rightarrow \infty)/\Omega_b}{0.02}.
\frac{\Omega_{d,0}(\Delta t \rightarrow \infty)}{0.02\cdot \Omega_b}
	\approx 1147\cdot \Omega_{d,0} (\Delta t \rightarrow \infty)
\end{equation}
or in absolute terms $Z \approx 22.9\Omega_{d,0}$.
At $z=5$ we have $\Omega_{d,0} = 2.3\cdot10^{-5} f_d$, which gives
$Z\approx 5.2\cdot10^{-4} f_d = 0.026 f_d\, Z_\odot$. %(by simply letting $\Delta t\rightarrow\infty$ in

There are not much metallicity data available for $z>5$. Metal poor stars in our galaxy are one point
of reference, absorption lines in the Ly$\alpha$ spectrum from quasars are another one.
The lowest metallicities found in stars in the Milky Way are $Z/Z_\odot \sim 0.01$,
\cite{2002A&A...390..187D}.
The Ly$\alpha$ forest suggests \cite[ figure 13]{2002AJ....123.2183S} that
$Z/Z_\odot \sim 0.003$ for $z \sim 4.5$ assuming that [Fe/H] $\approx$ log$(Z/Z_\odot)$
as suggested by \cite[ page 432]{2000ApJ...532..430V}.
This indicates that $f_d \sim 0.1$.
However, this might be lower than the actual value, cf. \cite[ figure 4]{1997ApJ...486..665P}.

In heavy stars, virtually all the helium is consumed, producing metals.
For simplicity (and lack of data), we assume that all the ejected metals clump to form dust,
$f_d \approx f_{eject}$.
This means that $f_d$ will almost entirely depend on the 
dust ejection rate in the supernova explosion. In
\cite{1998Natur.395..672I} a detected hypernova of mass $M\sim 14M_\odot$
seems to have $f_{eject}\gtrsim 0.7$.
Furthermore, according to a dust production model by \cite{2003astro.ph..7108N},
$f_d \approx 0.2-0.3$.
At the same time,
some of the stars will become black holes,
not ejecting any metals, \cite{2002ApJ...567..532H}, decreasing $f_d$.
Currently this decrease is largely unknown.

In summary, the mass fraction of the produced metals in the Pop III stars, having
become interstellar dust, should be around $f_d\sim0.1-0.3$.
In the following we use the more conservative $f_d = 0.1$, in agreement with
the Ly$\alpha$ forest measurements, unless otherwise stated.

\subsection{Dust Opacity}
\label{sec:DustOpacity}
With our model for the dust density evolution, we want to calculate the
opacity of the dust, as seen by the CMB. This will tell us how much
the CMB spectrum is altered by the passage through the dust.

The dust opacity is given by
\begin{eqnarray}
	\tau_{\nu}&=&c\int dz\frac{dt}{dz}\sigma_{\nu_e}n_d(z) \\
	&=& \frac{Q_0c}{\sqrt{\Omega_m}a_rH_0}\frac{3}{4}\frac{\rho_c}{\rho_g}
       \int dz\,\left(\frac{\nu}{\nu_r}\right)^{\beta_{\nu_e}}\Omega_{d,0}(z)
        (1+z)^{1/2+{\beta_{\nu_e}}},
\end{eqnarray}
where $\nu$ ($\nu_e$) is the observed (emitted) frequency and $\nu = \nu_e/(1+z)$.
The dust number density is $n_d(z) = (1+z)^3\times\rho_c\Omega_{d,0}(z)/m_g$ where
$m_g = \frac{4\pi a^{3}}{3}\rho_{g}$ is the grain mass.
We see (from $\tau \propto \Omega_{d,0}$)
that $\tau $ is proportional to the parameter $f_d$.

\begin{figure}
	\resizebox{\hsize}{!}{\includegraphics{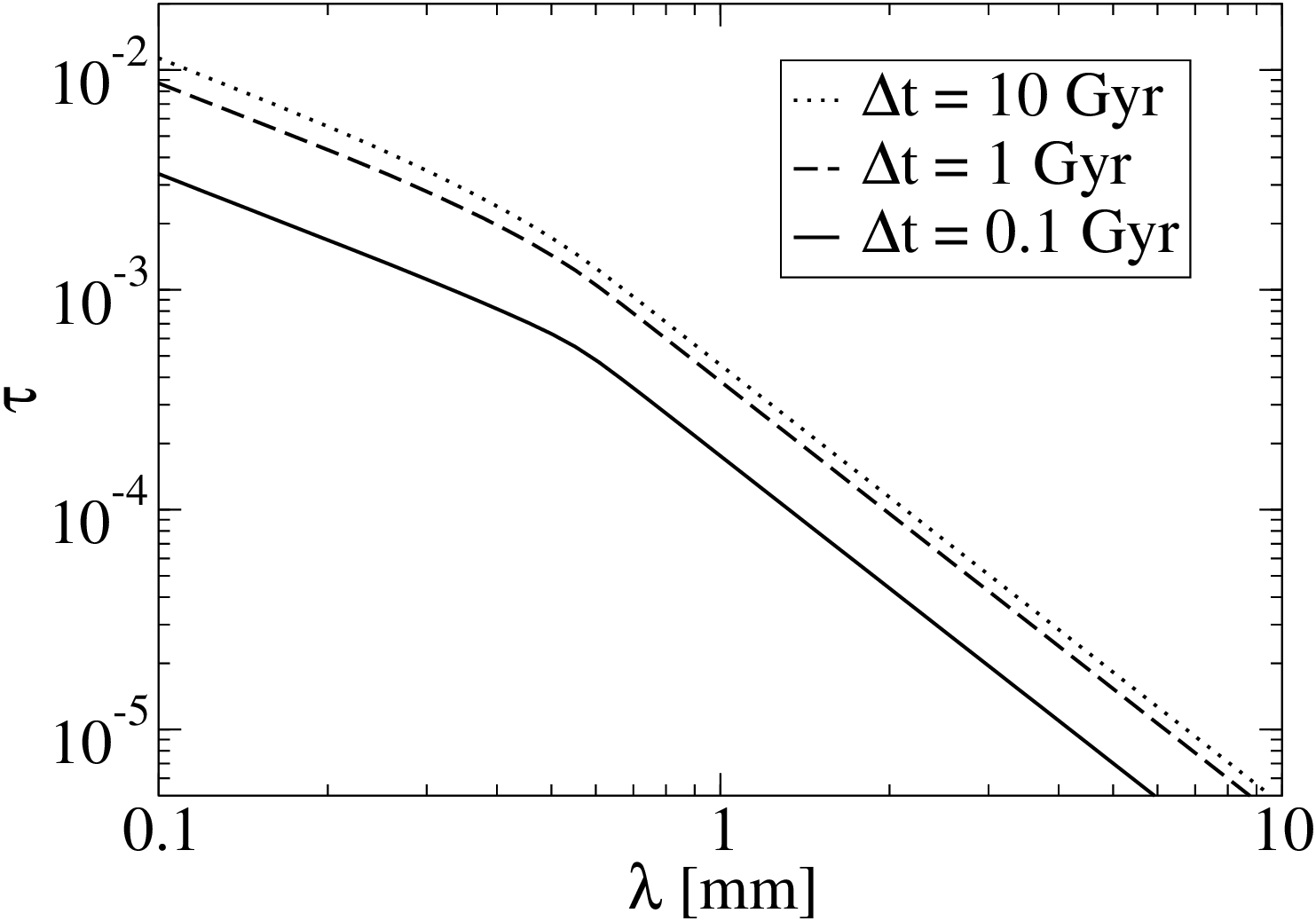}}
	\caption{Opacity $\tau$ with dust evolution taken into account. %The dashed line is included
	}
	\label{fig:tau_dust}
\end{figure}

\begin{figure}
\center
	\resizebox{\hsize}{!}{\includegraphics{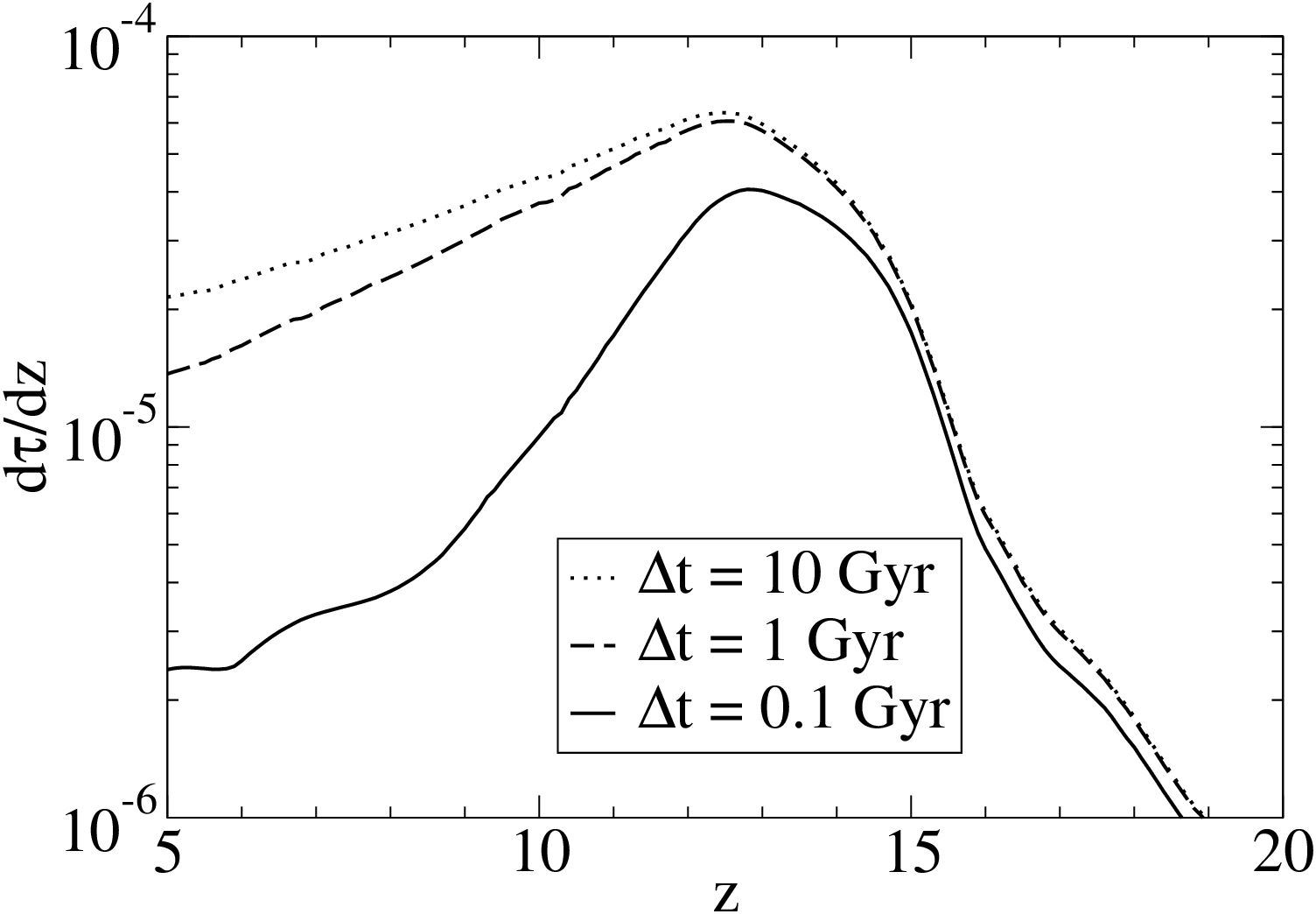}}
	\caption{The differential opacity $d\tau/dz$ at $\lambda = 1$ mm for different dust lifetimes.}
        \label{fig:dtau}
\end{figure}

The resulting opacity can be seen in Fig.~\ref{fig:tau_dust}.
We note that the opacity is small, $\tau \ll 1$.
The smooth knee is due to the change of $\beta$ at the redshifted $\nu_r$,
see Fig.~\ref{sec:DustEv}, but this is not in the spectral range of the CMB.
The differential opacity $d\tau/dz$ is plotted in Fig.~\ref{fig:dtau} for
$\lambda = 1$ mm.
We see that with a short dust lifetime, the dust density falls off
almost immediately (in terms of $z$). However, for longer lifetimes,
the early dust could still play a certain role for $z<3$. This could
eventually contribute to dimming of distant objects.
We also note the impact of the expansion of the universe in decreasing
the dust density and thus the opacity. This is why the increase in Fig.~\ref{fig:ng},
at $z\sim 5$, is not apparent in the opacity, Fig.~\ref{fig:tau_dust}.
The submillimetre effective dust opacity follows a $\nu^2$ emissivity law.


\subsection{Dust Temperature}
\label{sec:DustTemp}
In order to deduce the equilibrium temperature of the dust,
we write the balance between the
absorbed CMB, the absorbed starlight and the emitted IR light from the dust:

\begin{equation}
	P_d = P_* + P_{CMB}.
\label{eq:Pd-org}
\end{equation}
The powers $P_d$ and $P_{CMB}$ can be written as 
\begin{equation}
	P_{X} = 4\pi\int_0^\infty d\nu_e \,\sigma_{\nu_e} B_{\nu_e}(T_X),
\label{eq:P_X}
\end{equation}
where $B_\nu$ is a Planck blackbody spectrum and $X=\{CMB, d\}$.
In the wave-length range considered, the spectral index $\beta =2$.
Supposing that $\beta$ is constant, eq.~\ref{eq:Pd-org} can be solved for the dust temperature
analytically in the submm range:
\begin{equation}
	T_d^{4+\beta} = 
	T_{* eff}^{4+\beta}
	+T_{CMB}^{4+\beta},
\end{equation}
where the effective temperature is defined by
\begin{equation}
	T_{* eff}^{4+\beta} = \frac{P_*}
	  {8\pi^2 hc^{-2}(Q_0\cdot(a^3/a_r)\nu_r^{-\beta})(k_B/h)^{4+\beta}C_\beta}
\label{eq:Teff}
\end{equation}
and 
$	C_\beta = 
	\int_0^\infty dx x^{3+\beta}/(e^x-1) = 
	(\beta+3)!\sum_{k=1}^\infty k^{-(4+\beta)},$
such that $C_0 \approx 6.494$, $C_1 \approx 24.89$ and $C_2 \approx 122.1$.

However, in our calculations we use the exact eq. \ref{eq:Pd-org} and \ref{eq:P_X},
 while eq. \ref{eq:Teff} can be used as a cross-check.

The absorbed power density, $P_*$ from the radiation of Pop III stars peaks in the
UV-region and can be approximated by
\begin{equation}
	P_* = \sigma_{UV} u_*(z) c, 
\end{equation}
where $\sigma_{UV}$ is the dust-photon cross section in the UV region and the energy density is
\begin{equation}
	 u_*(z) = f_{esc}\int_{z_i}^{z} dz' \frac{dn_\gamma}{dz'}E_\gamma\left(\frac{1+z}{1+z'}\right)^4,
\end{equation}
where $f_{esc}$ is the escape fraction of photons from the star halos.
We neglect the loss of photons due to the reionization itself.
$E_\gamma\frac{1+z}{1+z'}$ is the effective energy of the photon emitted at $z'$
and then redshifted to $z$. 
According to \cite{Cen:2003ey}, $f_{esc} = 0.3$
gives an electron opacity $\tau_e \approx 0.13$ which
is within one standard deviation of the results by WMAP.
Hereafter, we adopt this value of $f_{esc}$.

\begin{figure}%[here!]
	\center
	\resizebox{\hsize}{!}{\includegraphics{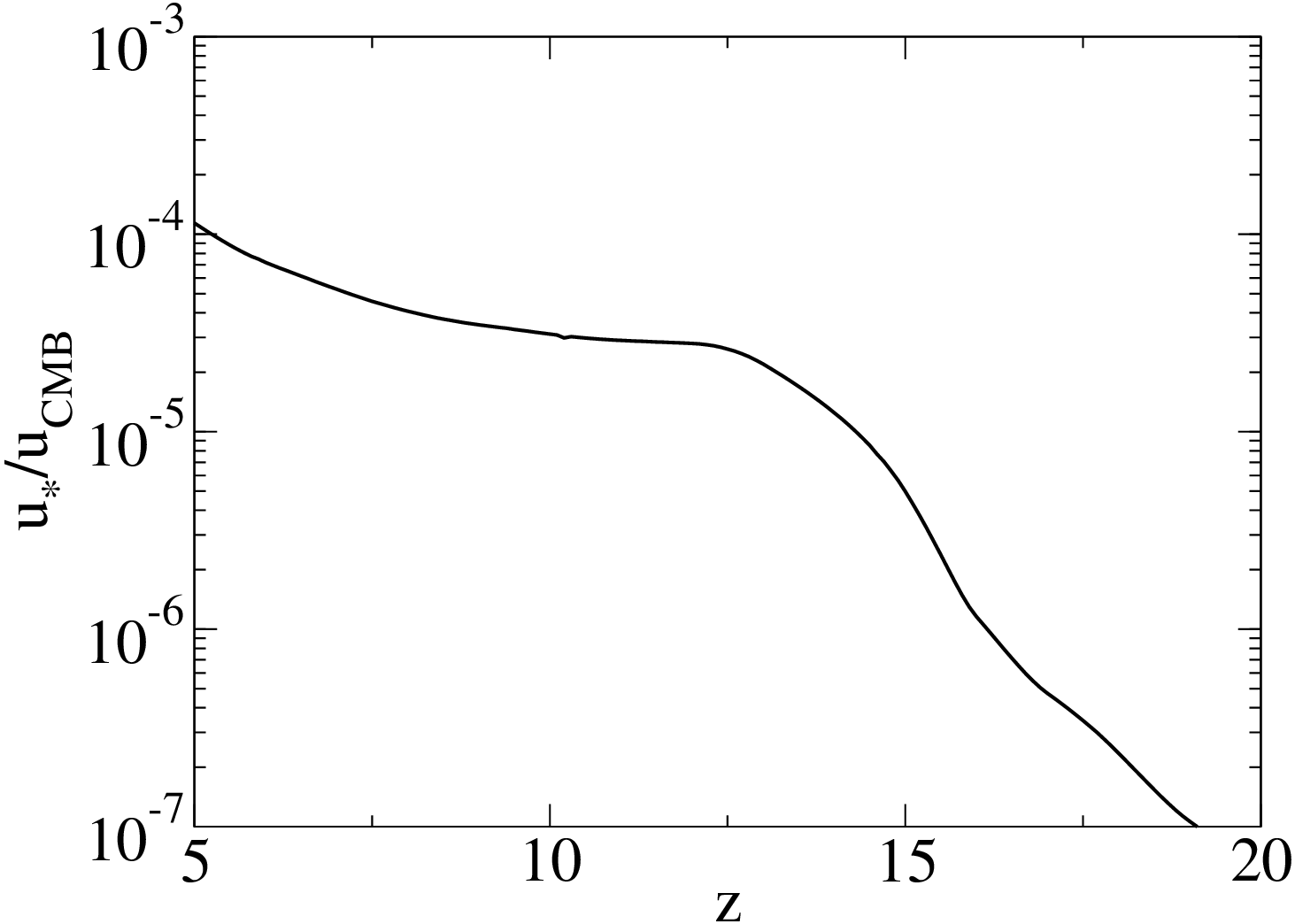}}
	\caption{Energy density of ionizing photons compared to $u_{CMB} = 4\sigma_ST_{CMB}^4/c$, where $\sigma_S$ is Stefan-Boltzmann's constant.}
	\label{fig:u_star}
\end{figure}

\begin{figure}%[here!]
	\center
	\resizebox{\hsize}{!}{\includegraphics{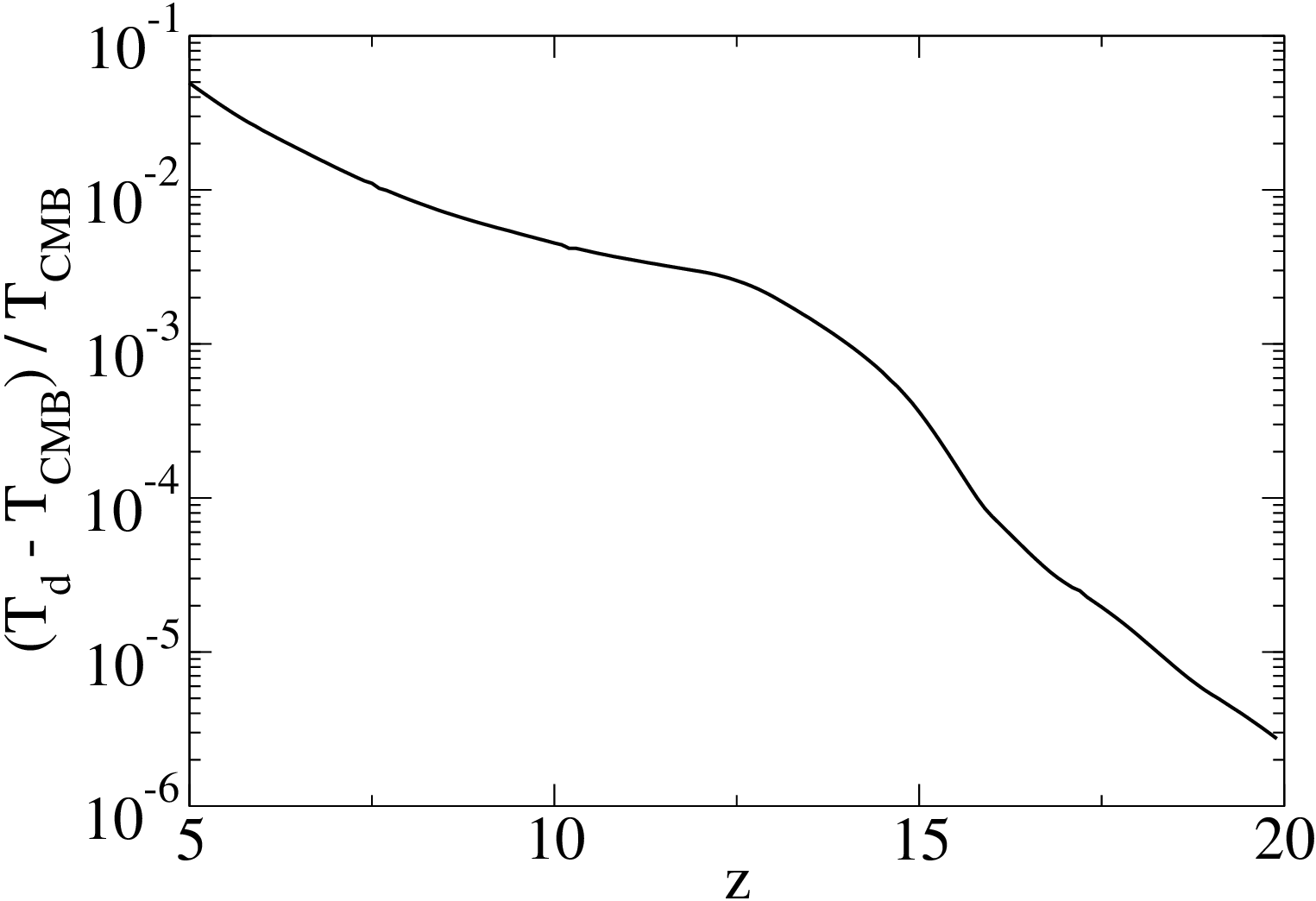}}
	\caption{The dust temperature is plotted against the CMB temperature
		with the relative quatity $(T_d-T_{CMB})/T_{CMB}$.
	}
	\label{fig:Td}
\end{figure}

The energy density of the ionizing photons are compared to the CMB in Fig.~\ref{fig:u_star}.
The star energy density is much less
than the CMB energy density at this epoch, and the curve resembles the accumulated
photons in Fig.~\ref{fig:ng}.
Hence, the dust temperature closely follows the CMB temperature, see Fig.~\ref{fig:Td} and
Eq.~\ref{eq:P_X}.

%}
%

\subsection{Observed Intensity}
Now we proceed to compute the average intensity (monopole term) of the
submm and microwave background which is made of the CMB and early dust emission.
The simple radiative transfer of the CMB through the uniform dust screen
yields the following observed intensity:
\begin{equation}
	i_\nu = e^{-\tau_\nu}\left[ B_\nu(T_{CMB}) + \int_0^{\tau_\nu}
	   e^{\tau_e} \frac{B_{\nu_e}(T_d(z))}{(1+z)^{3}}\,d\tau_e \right].
\end{equation}
From Fig.~\ref{fig:tau_dust} and \ref{fig:Td}, we see that the opacity is small,
($\tau \ll 1$) and the dust temperature is
only slightly higher than the CMB temperature ($T_d \gtrsim T_{CMB}$). This gives
the following formula for the excess intensity relative to the unperturbed CMB:
\begin{eqnarray}
	\Delta i_{\nu} &\equiv& i_{\nu} - B_{\nu}(T_{CMB}) \nonumber\\
	&=&\left. T_{CMB}\frac{dB_\nu}{dT}\right|_{T=T_{CMB}}
	\int_0^{\tau_\nu} \frac{ T_d(z) - T_{CMB}(z)}{T_{CMB}(z)} d\tau_e,
\label{eq:delta_i}
\end{eqnarray}
where $T_{CMB}$ is the CMB temperature today. % and $\Delta T(z) = T_d(z) - T_{CMB}(z)$.
The integrant is plotted in Fig.~\ref{fig:Td}.
We note that a new component is added to the primary CMB spectrum. Eq.~\ref{eq:delta_i}
tells us that it has a specific spectrum which is the product of a 2.725 K blackbody
temperature fluctuation spectrum (like primary anisotropies) and a $\nu^2$ power law
(from $d\tau_e$).
This effect is mostly visible in the submm range and has a minor contribution
in the radio domain.

\begin{figure}
	\resizebox{\hsize}{!}{\includegraphics{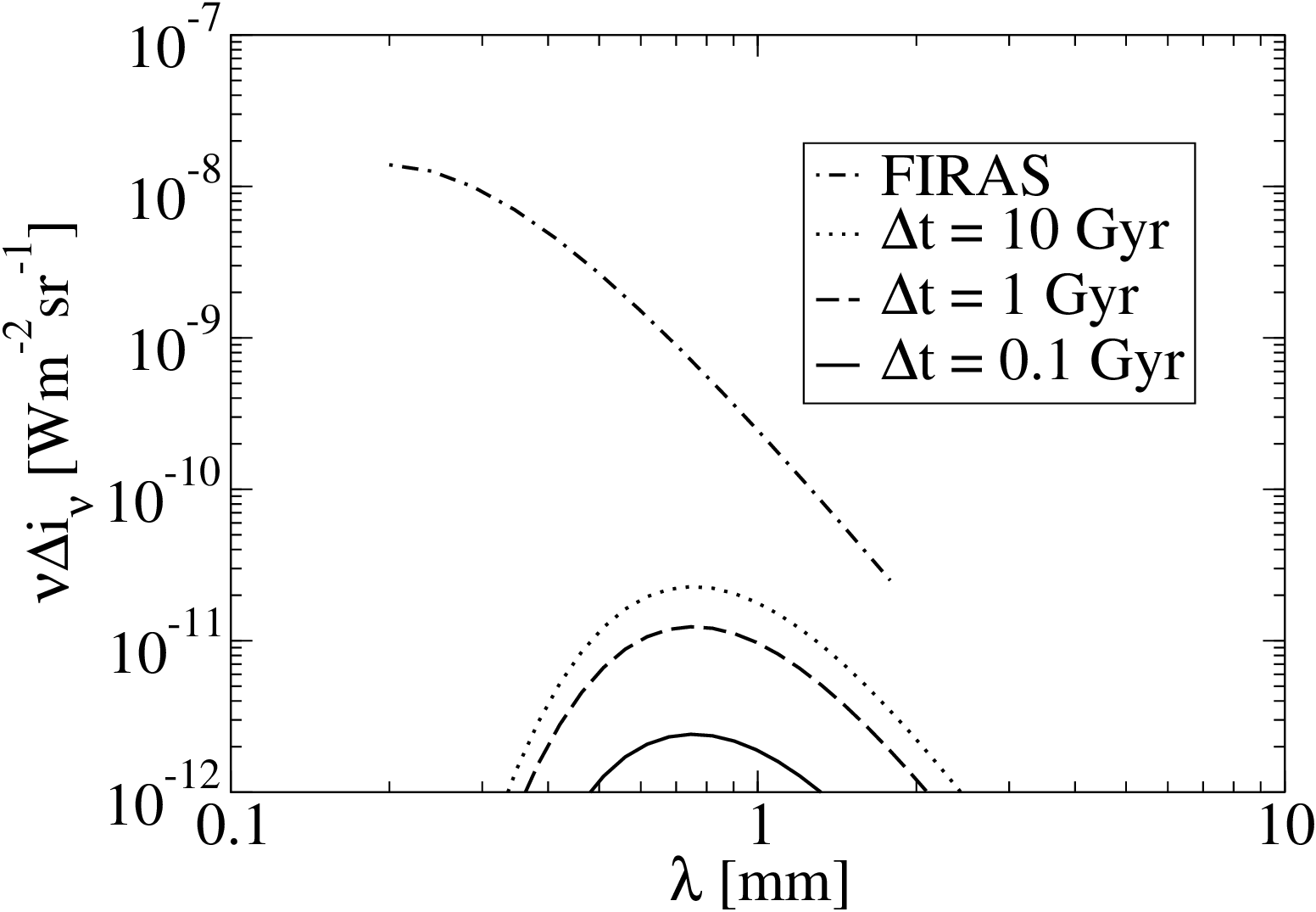}}
	\caption{Comparison of the modeled intensity for the early dust
	emission in excess of the CMB
	with the observed FIRAS spectrum (dashed red curve)
	of the cosmic far IR background
	as detailed by \cite{1999A&A...344..322L}.}
        \label{fig:FIRAS}
\end{figure}

In Fig.~\ref{fig:FIRAS}, the excess intensity is plotted along with the extragalactic
background measured by FIRAS,
\cite{1996A&A...308L...5P,1998ApJ...508..123F,1999A&A...344..322L}.
Depending on the dust destruction rate (parametrized by the dust lifetime $\Delta t$),
the computed early dust background can
be an important part of the observed background from 400 $\mu$m up to the
mm wave-length. The exact position of $\lambda_r$ will only slightly displace
the spectrum, leaving the magnitude unchanged.
Most of the far IR background can now be explained by
a population of $z=0$ to $z=3$ luminous IR galaxies, \cite{2000A&A...360....1G}.
A fraction of the submillimetre part of this background could arise
from larger redshift dust emission as suggested by Fig.~\ref{fig:FIRAS}.

In order to check our results, we calculate the co-moving luminosity density of the dust
in the submm region and compare it with \cite[ figure 4]{2000A&A...360....1G}.
We find them compatible.

\subsection{Discussion}

Just like the Thomson scattering during reionization,
early dust will also tend to erase the primordial anisotropies in the CMB.
However, due to the much smaller dust opacity (compare $\tau_d(1mm) \lesssim 10^{-3}$ and $\tau_e = 0.17$),
this effect will be negligible.

The early dust will also introduce a new type of
secondary anisotropies with a typical size of a dark matter filament.
Here, we only estimate the order of magnitude of this effect.
If the co-moving size of the dark matter filament is $L$,
the angular size is $3\cdot(L/5$ Mpc) arcminutes at $z=10$
which corresponds to multipole number $\ell \sim 4000\cdot(L/5$ Mpc). Fortunately,
this region in $\ell$-space
does not contain any primordial fluctuations because of the Silk damping.
However, there are other foregrounds in the
same region, see \cite{2000A&A...357....1A}.
If we suppose a contrast of 10\% in the
dust intensity between dark matter filaments and the void, we obtain
values of $\Delta T/T \approx 3\times 10^{-7}$ (for $\lambda=1$ mm, $f_d=0.1$ and $\Delta t = 1$ Gyr).
These anisotropies, pending more accurate 
calculations, clearly are in the range of expected arcminute 
secondary anisotropies from other effects. They could be detected by Planck HFI (High Frequency Instrument),
\cite{2003astro.ph..8075L} and FIRAS--II type of instrument, \cite{2002ApJ...581..817F}.

The results of these calculations depend only very weakly on the precise 
dust model assumptions.
We have also tried a different (but similar) shape of the ionizing
photon production, Fig.~\ref{fig:ng}, and found that the results do
not vary significantly.

Very little is known about the universe during the reionization epoch. Nevertheless,
there are several parameters that could be calculated more accurately.

The two most important parameters in the present model are the
dust lifetime, $\Delta t$ and the mass fraction
of the produced metals that are ejected as interstellar dust, $f_d$. 
The dust lifetime could be determined more precisely by making 3D simulations
of the dust production in combination with structure formation. The simulations would also
give the inhomogeneous dust density evolution. The result would be a better
estimate of the aforementioned secondary anisotropies caused by the variations in the dust opacity.
A more refined dust grain model, using e.g. a distribution of grain sizes
would also be more realistic. If the dust is long-lived, it could also
have a certain impact on measurements in the optical and UV region.
Finally, we note that most of the results are proportional to
the dust density and thus to $f_d$. To evaluate $f_d$ more precisely,
we need a better understanding of the typical properties of the
first generation of stars, see section~\ref{sec:Met}, which is
currently much debated.

\section{Conclusions}
We have shown that the radiation from early dust, produced and heated by Pop III stars,
contributes to the extragalactic submillimetre background within the limits set by FIRAS.
It may not be detected by the present generation of instruments
but future experiments such as Planck HFI and FIRAS-II should be able
to measure it, by using its specific predicted spectral
signature. This high--redshift dust, contemporary to the reionization,
should show up as small--scale anisotropies when observed by sensitive
submillimetre instruments. These anisotropies are in the same range as other
small--scale anisotropy effects.

\bibliographystyle{aa} % style aa.bst

\bibliography{0448bib}        % 0448bib.bib is the name of our database

\begin{thebibliography}{24}
\expandafter\ifx\csname natexlab\endcsname\relax\def\natexlab#1{#1}\fi

\bibitem[{{Aghanim} {et~al.}(2000){Aghanim}, {Balland}, \&
  {Silk}}]{2000A&A...357....1A}
{Aghanim}, N., {Balland}, C., \& {Silk}, J. 2000, \aap, 357, 1

\bibitem[{{Boulanger} {et~al.}(1996){Boulanger}, {Abergel}, {Bernard},
  {Burton}, {Desert}, {Hartmann}, {Lagache}, \& {Puget}}]{1996A&A...312..256B}
{Boulanger}, F., {Abergel}, A., {Bernard}, J.-P., {et~al.} 1996, \aap, 312, 256

\bibitem[{{{Cen}, R.}(2002)}]{Cen:2002zc}
{{Cen}, R.} 2002, \apj, 591, 12

\bibitem[{{{Cen}, R.}(2003)}]{Cen:2003ey}
{{Cen}, R.} 2003, \apjl, 591, L5

\bibitem[{{Depagne} {et~al.}(2002){Depagne}, {Hill}, {Spite}, {Spite}, {Plez},
  {Beers}, {Barbuy}, {Cayrel}, {Andersen}, {Bonifacio}, {Fran{\c c}ois},
  {Nordstr{\" o}m}, \& {Primas}}]{2002A&A...390..187D}
{Depagne}, E., {Hill}, V., {Spite}, M., {et~al.} 2002, \aap, 390, 187

\bibitem[{{Desert} {et~al.}(1990){Desert}, {Boulanger}, \&
  {Puget}}]{1990A&A...237..215D}
{Desert}, F.-X., {Boulanger}, F., \& {Puget}, J.~L. 1990, \aap, 237, 215

\bibitem[{{Draine}(1990)}]{1990eism.conf..193D}
{Draine}, B.~T. 1990, in ASP Conf. Ser. 12: The Evolution of the Interstellar
  Medium, 193--205

\bibitem[{{Draine} \& {Salpeter}(1979)}]{1979ApJ...231..438D}
{Draine}, B.~T. \& {Salpeter}, E.~E. 1979, \apj, 231, 438

\bibitem[{{Fixsen} {et~al.}(1998){Fixsen}, {Dwek}, {Mather}, {Bennett}, \&
  {Shafer}}]{1998ApJ...508..123F}
{Fixsen}, D.~J., {Dwek}, E., {Mather}, J.~C., {Bennett}, C.~L., \& {Shafer},
  R.~A. 1998, \apj, 508, 123

\bibitem[{{Fixsen} \& {Mather}(2002)}]{2002ApJ...581..817F}
{Fixsen}, D.~J. \& {Mather}, J.~C. 2002, \apj, 581, 817

\bibitem[{{Gispert} {et~al.}(2000){Gispert}, {Lagache}, \&
  {Puget}}]{2000A&A...360....1G}
{Gispert}, R., {Lagache}, G., \& {Puget}, J.~L. 2000, \aap, 360, 1

\bibitem[{{Heger} \& {Woosley}(2002)}]{2002ApJ...567..532H}
{Heger}, A. \& {Woosley}, S.~E. 2002, \apj, 567, 532

\bibitem[{{Iwamoto} {et~al.}(1998){Iwamoto}, {Mazzali}, {Nomoto}, {Umeda},
  {Nakamura}, {Patat}, {Danziger}, {Young}, {Suzuki}, {Shigeyama},
  {Augusteijn}, {Doublier}, {Gonzalez}, {Boehnhardt}, {Brewer}, {Hainaut},
  {Lidman}, {Leibundgut}, {Cappellaro}, {Turatto}, {Galama}, {Vreeswijk},
  {Kouveliotou}, {van Paradijs}, {Pian}, {Palazzi}, \&
  {Frontera}}]{1998Natur.395..672I}
{Iwamoto}, K., {Mazzali}, P.~A., {Nomoto}, K., {et~al.} 1998, \nat, 395, 672

\bibitem[{{Kogut} {et~al.}(2003){Kogut}, {Spergel}, {Barnes}, {Bennett},
  {Halpern}, {Hinshaw}, {Jarosik}, {Limon}, {Meyer}, {Page}, {Tucker},
  {Wollack}, \& {Wright}}]{2003ApJS..148..161K}
{Kogut}, A., {Spergel}, D.~N., {Barnes}, C., {et~al.} 2003, \apjs, 148, 161

\bibitem[{{Lagache} {et~al.}(1999){Lagache}, {Abergel}, {Boulanger}, {D{\'
  e}sert}, \& {Puget}}]{1999A&A...344..322L}
{Lagache}, G., {Abergel}, A., {Boulanger}, F., {D{\' e}sert}, F.~X., \&
  {Puget}, J.-L. 1999, \aap, 344, 322

\bibitem[{{Lamarre} {et~al.}(2003){Lamarre}, {Puget}, {Bouchet}, {Ade},
  {Benoit}, {Bernard}, {Bock}, {De Bernardis}, {Charra}, {Couchot},
  {Delabrouille}, {Efstathiou}, {Giard}, {Guyot}, {Lange}, {Maffei}, {Murphy},
  {Pajot}, {Piat}, {Ristorcelli}, {Santos}, {Sudiwala}, {Sygnet}, {Torre},
  {Yurchenko}, \& {Yvon}}]{2003astro.ph..8075L}
{Lamarre}, J.~M., {Puget}, J.~L., {Bouchet}, F., {et~al.} 2003, ArXiv
  Astrophysics e-prints, astro-ph/0308075

\bibitem[{{Nozawa} {et~al.}(2003){Nozawa}, {Kozasa}, {Umeda}, {Maeda}, \&
  {Nomoto}}]{2003astro.ph..7108N}
{Nozawa}, T., {Kozasa}, T., {Umeda}, H., {Maeda}, K., \& {Nomoto}, K. 2003,
  ArXiv Astrophysics e-prints, astro-ph/0307108

\bibitem[{{Pei} {et~al.}(1999){Pei}, {Fall}, \& {Hauser}}]{1999ApJ...522..604P}
{Pei}, Y.~C., {Fall}, S.~M., \& {Hauser}, M.~G. 1999, \apj, 522, 604

\bibitem[{{Pettini} {et~al.}(1997){Pettini}, {Smith}, {King}, \&
  {Hunstead}}]{1997ApJ...486..665P}
{Pettini}, M., {Smith}, L.~J., {King}, D.~L., \& {Hunstead}, R.~W. 1997, \apj,
  486, 665

\bibitem[{{Puget} {et~al.}(1996){Puget}, {Abergel}, {Bernard}, {Boulanger},
  {Burton}, {Desert}, \& {Hartmann}}]{1996A&A...308L...5P}
{Puget}, J.-L., {Abergel}, A., {Bernard}, J.-P., {et~al.} 1996, \aap, 308, L5+

\bibitem[{{Shioya} {et~al.}(2002){Shioya}, {Taniguchi}, {Murayama}, {Nishiura},
  {Nagao}, \& {Kakazu}}]{2002ApJ...576...36S}
{Shioya}, Y., {Taniguchi}, Y., {Murayama}, T., {et~al.} 2002, \apj, 576, 36

\bibitem[{{Songaila} \& {Cowie}(2002)}]{2002AJ....123.2183S}
{Songaila}, A. \& {Cowie}, L.~L. 2002, \aj, 123, 2183

\bibitem[{{Spergel} {et~al.}(2003){Spergel}, {Verde}, {Peiris}, {Komatsu},
  {Nolta}, {Bennett}, {Halpern}, {Hinshaw}, {Jarosik}, {Kogut}, {Limon},
  {Meyer}, {Page}, {Tucker}, {Weiland}, {Wollack}, \&
  {Wright}}]{2003ApJS..148..175S}
{Spergel}, D.~N., {Verde}, L., {Peiris}, H.~V., {et~al.} 2003, \apjs, 148, 175

\bibitem[{{VandenBerg} {et~al.}(2000){VandenBerg}, {Swenson}, {Rogers},
  {Iglesias}, \& {Alexander}}]{2000ApJ...532..430V}
{VandenBerg}, D.~A., {Swenson}, F.~J., {Rogers}, F.~J., {Iglesias}, C.~A., \&
  {Alexander}, D.~R. 2000, \apj, 532, 430

\end{thebibliography}

\end{document}